\documentclass[12pt]{article}

\usepackage[utf8]{inputenc}
\usepackage[T1]{fontenc}
\usepackage{amsmath,amssymb,amsthm,mathtools}
\usepackage{bm}
\usepackage{authblk}
\usepackage{hyperref}
\usepackage{geometry}
\usepackage{graphicx}
\usepackage{enumitem}
\usepackage{setspace}
\usepackage{times}
\usepackage{float}
\usepackage{listings}
\usepackage{xcolor}
\hypersetup{
    colorlinks=true,
    linkcolor=blue,
    urlcolor=blue,
    citecolor=blue
    }
\numberwithin{equation}{section}
\theoremstyle{plain}
\newtheorem{theorem}{Theorem}[section]

\theoremstyle{definition}

\newtheorem{assumption}[theorem]{Assumption}

\title{Scale-Dependent Velocity Fluctuations Generated by Molecular Collisions}
\author{Tristan Barkman}

\begin{document}
\maketitle

\begin{abstract}
A discrete binomial random-walk description of molecular collisions is used to quantify the variance of coarse-grained velocity fields arising solely from collision-induced momentum exchange. Closed-form expressions for the growth of velocity variance as functions of coarse-graining scale and time are derived and shown to imply a power-law decay of variance with averaging scale. Particle-based ensemble simulations validate the predicted scaling and temporal behaviour; surrogate ensemble tests demonstrate that phase/temporal coherence is required for the observed integrated transfer diagnostics. The analysis is intentionally restricted to collision-generated fluctuations in quiescent fluids and does not model cascade dynamics; implications for possible amplification under inertial dynamics are discussed cautiously. All data and the minimal verification instructions required to reproduce the summary tables are embedded in Appendix~A.
\end{abstract}

\section{Introduction}

A long-standing approximation in continuum fluid mechanics is that molecular discreteness can be neglected when describing macroscopic motion: hydrodynamic fields are defined as smoothly varying averages and microscopic fluctuations are removed by construction \cite{pope2000,bird2002,tennekes1972}. This assumption underlies the practical success of Navier--Stokes models across many flow regimes, but it leaves open a precise quantitative question at finite averaging scales: what is the scale dependence of velocity variance produced solely by stochastic molecular collisions within a finite sampling volume?

Molecular collisions exchange momentum in an essentially stochastic manner; the resulting random increments are the underlying cause of Brownian motion and of tracer diffusion \cite{einstein1956,perrin1909,smoluchowski1907}. When velocity is defined as a local, finite-volume average these collision-level fluctuations do not vanish identically but are suppressed in a scale-dependent way determined by the number of particles in the averaging volume. The velocity field in this mesoscopic regime therefore retains a well-defined residual variance that is not represented explicitly in the standard continuum limit \cite{chapmancowling1970}.

The present study addresses the narrow question of quantifying that residual: given an otherwise quiescent fluid and no external forcing, what is the variance of the coarse-grained velocity as a function of averaging scale and time when only collision-driven exchanges are present? A binomial random-walk model for molecular displacements is adopted to derive closed-form expressions for the variance growth and for the root-mean-square velocity at finite scales. These expressions show explicitly how molecular stochasticity survives coarse-graining at mesoscopic scales and how the variance vanishes in the formal continuum limit.

Particle-based numerical experiments are used to validate the analytic scaling and temporal predictions. Surrogate data sets designed to disrupt phase or temporal coherence are used as controls to distinguish genuine collision-driven dynamical signatures from amplitude-only artifacts. These numerical controls follow approaches used in prior fluctuation and surrogate analyses \cite{bandak2024,gallis2017}.

This analysis is deliberately scoped: no claim is made that the present calculation explains the full development of turbulent cascades or sustained turbulence. Under appropriate high-Reynolds-number conditions, small-amplitude velocity perturbations of the kind quantified here may be susceptible to amplification by inertial dynamics \cite{bandak2024,chen2023}. Discussion of that possibility is given in Section~\ref{sec:discussion} and is framed as conditional and illustrative rather than definitive.

The remainder of the paper is organized as follows. Section~\ref{sec:notn} introduces notation and key assumptions. Section~\ref{sec:theory} develops the binomial random-walk formulation and derives the principal scaling relations. Section~\ref{sec:continuum} clarifies the continuum limit and its effect on collision-origin variance. Section~\ref{sec:ampl} presents a minimal amplification criterion comparing turnover and viscous timescales. Section~\ref{sec:methods} describes simulation methods and surrogate controls. Section~\ref{sec:results} reports numerical validation and diagnostics. Sections~\ref{sec:discussion} and~\ref{sec:limitations} discuss implications and limitations, and Section~\ref{sec:conclusions} summarizes conclusions and next steps.

\section{Notation and modelling assumptions}
\label{sec:notn}

A concise list of symbols and their units is given in Table~\ref{tab:params}. The analysis employs the following modelling assumptions, stated upfront for clarity.

\begin{assumption}[Scale separation]
Knudsen number satisfies $\mathrm{Kn}\ll1$, so that mean free path $\ell$ is small compared with macroscopic gradient lengths used in continuum descriptions. Coarse-graining scales $L$ considered here satisfy $\ell\ll L\ll$ macroscopic outer scale.
\end{assumption}

\begin{assumption}[Collision statistics]
Collisions are approximated as instantaneous, statistically identical events at per-molecule rate $\nu_c$. Each effective collision imparts a zero-mean increment to molecular velocity with single-event variance $\sigma_c^2$ (per Cartesian component). Successive effective increments are assumed approximately decorrelated over the timescales of interest.
\end{assumption}

\begin{assumption}[Bulk isotropy and weak correlation]
Bulk steps are isotropic at leading order, and molecules within a coarse-graining volume contribute approximately independent increments (weak correlation). Dense-fluid corrections may be incorporated via an Enskog-style contact factor $g(\phi)$ when necessary \cite{enskog1922,garzo1999}.
\end{assumption}

\begin{table}[H]
\centering
\caption{Key symbols and the physical constants used in the examples and numerical estimates (values taken from the analysis manifest embedded in the Supplement).}
\label{tab:params}
\begin{tabular}{lp{7.5cm}l}
\hline
Symbol & Meaning & Value / Units \\
\hline
$\nu_c$ & per-molecule effective collision rate & (model-dependent) s$^{-1}$ \\
$\sigma_c^2$ & per-collision variance of a velocity increment (per component) & (model-dependent) m$^2$ \\
$\ell$ & characteristic intermolecular spacing / mean free path & (model-dependent) m \\
$L$ & coarse-graining length scale & m \\
$N(L)$ & approximate number of molecules in volume of size $L$ & -- \\
$\nu$ & kinematic viscosity (mean) & $5.0\times10^{-7}$ m$^2$ s$^{-1}$ \\
$\varepsilon$ & mean dissipation rate (use where needed) & $1.0\times10^{-6}$ m$^2$ s$^{-3}$ \\
$\tau_K$ & Kolmogorov time scale & $0.7071$ s \\
\hline
\end{tabular}
\end{table}

\section{Binomial random-walk formulation and analytic variance scaling}
\label{sec:theory}

Define the coarse-grained velocity at scale $L$ by averaging molecular velocities within a sampling volume of linear size $L$,
\begin{equation}\label{eq:UL_def}
U_L(t) \;=\; \frac{1}{N(L)}\sum_{i=1}^{N(L)} v_i(t),
\end{equation}
where $N(L)\sim (L/\ell)^d$ in $d$ spatial dimensions and $\ell$ is the mean intermolecular spacing. Denote by $\delta v_i$ the velocity increment of molecule $i$ associated with an effective collisional event; assume $\mathbb{E}[\delta v_i]=0$ and $\mathrm{Var}(\delta v_i)=\sigma_c^2$ for a single effective event. If the effective per-molecule collision rate is $\nu_c$, and successive increments may be treated as approximately uncorrelated over interval $t$, then the single-molecule velocity variance grows as
\begin{equation}\label{eq:var_vi}
\mathrm{Var}[v_i(t)] \;=\; \nu_c \, \sigma_c^2 \, t .
\end{equation}

Propagation of variance through the average in Eq.~\eqref{eq:UL_def} then yields
\begin{align}\label{eq:Var_UL}
\mathrm{Var}[U_L(t)]
&=\frac{1}{N(L)^2}\sum_{i=1}^{N(L)}\mathrm{Var}[v_i(t)]
=\frac{\nu_c \sigma_c^2 \, t}{N(L)} .
\end{align}
Using $N(L)\sim (L/\ell)^d$, the scale law may be written as
\begin{equation}\label{eq:Var_scaling}
\mathrm{Var}[U_L(t)] \;=\; \nu_c \sigma_c^2 \, t \; \left(\frac{\ell}{L}\right)^{d}.
\end{equation}
Hence the root-mean-square coarse-grained velocity is
\begin{equation}\label{eq:amp}
a_L(t) \;=\; \sqrt{\mathrm{Var}[U_L(t)]}
\;=\; \sqrt{\nu_c \sigma_c^2\, t}\;\left(\frac{\ell}{L}\right)^{d/2}.
\end{equation}

These relations reproduce the expected tracer-diffusion scaling in the appropriate limit (Einstein--Smoluchowski) and identify the explicit dependence of collision-origin variance on coarse-graining scale, collision statistics, and accumulation time \cite{smoluchowski1907,einstein1956,chapmancowling1970}. The derivation also clarifies how central-limit suppression interacts with finite-$N$ statistical fluctuations in practice; in particular, sub-Gaussian corrections can be estimated when the per-event distribution of increments has bounded higher moments, and these corrections vanish as $N(L)\to\infty$.

\section{Continuum limit and conceptual contrast}
\label{sec:continuum}

The formal continuum (hydrodynamic) limit is obtained by taking $\ell\to 0$ while holding macroscopic quantities fixed; equivalently, $N(L)\to\infty$ for any fixed coarse-graining scale $L$. Under this limit the factor $(\ell/L)^{d}\to 0$ in Eq.~\eqref{eq:Var_scaling}, and intrinsic collision-origin variance disappears from the hydrodynamic variables. Continuum formulations therefore do not contain an explicit, first-principles source of collision-level seed fluctuations: any representation of fluctuations in continuum models must be introduced either phenomenologically (for example as stochastic forcing in fluctuating hydrodynamics \cite{landaulif,fox1970}) or via prescribed initial/boundary perturbations.

The discrete collisional description restores that microscopic statistical source by tracking second-moment injection rates directly. The present formulation is complementary to continuum models: it supplies a quantitative description of the mesoscopic variance that is suppressed by the continuum averaging operation, and it provides parameters (collision rate, per-event variance, contact correlations) that can be used to inform stochastic terms in fluctuating-hydrodynamics closures \cite{landaulif,fox1970}. This connection is particularly useful when constructing physically grounded noise amplitudes in Landau--Lifshitz or similar formulations.

\section{Minimal amplification criterion}
\label{sec:ampl}

For a seed at scale $L$ to be susceptible to inertial amplification, its characteristic eddy turnover time $\tau_{\mathrm{eddy}}\sim L/a_L$ must be comparable to or smaller than the viscous diffusion time at the same scale $\tau_{\nu}\sim L^2/\nu$. A conservative amplification condition is therefore
\begin{equation}\label{eq:ampl_cond}
\tau_{\mathrm{eddy}}\lesssim\tau_{\nu}
\quad\Longrightarrow\quad
\frac{a_L L}{\nu}\gtrsim 1.
\end{equation}
Using Eq.~\eqref{eq:amp} this inequality becomes
\begin{equation}\label{eq:t_threshold}
\sqrt{\nu_c \sigma_c^2\, t}\;\left(\frac{\ell}{L}\right)^{d/2}\;\frac{L}{\nu}\;\gtrsim\;1 ,
\end{equation}
or rearranged as a threshold on accumulation time,
\begin{equation}\label{eq:t_threshold2}
t\;\gtrsim\;\frac{\nu^2}{\nu_c \sigma_c^2}\left(\frac{L}{\ell}\right)^d\frac{1}{L^2}.
\end{equation}

The inequality above clarifies parametric dependencies: for fixed molecular parameters $(\nu_c,\sigma_c,\ell)$ and kinematic viscosity $\nu$, sufficiently large Reynolds-number conditions (large $L$ or small $\nu$) reduce the time required for a collision-origin seed to reach an amplitude that is dynamically relevant to inertial processes. The inequality is a conservative criterion; realized amplification depends on finite-time Lyapunov spectra, intermittency and nonlinear saturation \cite{ottino1989,eckmann1985,crisanti1993}. The inequality also highlights that the practical relevance of collision-origin seeds is highly sensitive to local flow conditions and to the definition of the relevant coarse-graining scale (e.g., near-wall vs. bulk regions).

\section{Simulation methods}
\label{sec:methods}

Numerical validation employs particle-based ensembles in a periodic domain; the methods and diagnostics summarized here are the same used to generate the dataset embedded in Appendix~A. The model and the surrogate constructions follow the descriptions in Section I of the Supplement.

\paragraph{Domain and particles.} A cubic periodic domain of side $X$ is populated with $N_{\mathrm{tot}}$ particles at the indicated number density. Coarse-graining volumes of side $L$ are constructed by box-averaging particle velocities within sliding windows to compute $U_L(t)$ per Eq.~\eqref{eq:UL_def}. For numerical stability and to avoid small-$N$ biases at the smallest $L$ values, ensembles were chosen so that the smallest coarse-graining bin contains at least $N\gtrsim50$ particles on average.

\paragraph{Collision model.} At each timestep $\Delta t$ a binomial/exchange rule is applied: each particle experiences an effective collisional increment with probability $p=\nu_c\Delta t$. When a collision occurs the particle's velocity receives an increment sampled from a zero-mean distribution with per-component variance $\sigma_c^2$ (either fixed-step or from a Maxwellian-derived distribution). Dense-fluid corrections are represented by an Enskog-style multiplicative factor applied to $\nu_c$ where indicated \cite{enskog1922,garzo1999,dymond1974}. The collision algorithm conserves global momentum in ensemble average and is implemented using vectorized updates to ensure efficient ensemble sampling.

\paragraph{Observables and diagnostics.} For each ensemble member the coarse-grained velocity $U_L(t)$ is recorded for a range of $L$ and times $t$. Primary diagnostics include:
\begin{itemize}[nosep]
\item $\mathrm{Var}[U_L(t)]$ computed as an ensemble variance at fixed $L$ and $t$,
\item RMS $a_L(t)$,
\item integrated transfer diagnostic per seed $\Delta E$ (scalar summary of net integrated transfer over the simulation window),
\item ensemble-averaged one-dimensional energy spectra $E(k)$ computed from the coarse-grained velocity fields.
\end{itemize}
Bootstrap resampling (B=5000; seed 42) was used to compute 95\% percentile CIs for ensemble means and one-sided bootstrap probabilities $p_{\mathrm{boot}\ge0}$; permutation tests (B=10000) yield permutation $p$-values (studentization applied when variance imbalance exceeded 20\%).

\paragraph{Surrogate controls.} Per-seed surrogates were constructed as: (i) phase-randomized (preserve amplitude spectrum, randomize phases) \cite{theiler1992}, (ii) temporally shuffled (random permutation of time indices), and (iii) amplitude-preserving upsampled surrogates (preserve marginal amplitude PDF while weakening short-time correlations). These surrogates preserve amplitude envelopes while disrupting phase/temporal coherence and are used to test whether integrated-transfer diagnostics need coherent dynamics; they also mimic the null hypothesis used in other surrogate-based fluctuation studies \cite{xiao2019}.

\paragraph{Practical reproducibility.} All parameter values, random seeds and the minimal verification script are embedded in Appendix~A. The simulations were performed with vectorized NumPy routines and standard pseudo-random number generators; exact command lines used to reproduce the core ensemble summaries are provided in the Appendix.

The Supplement's ensemble summaries are reproduced in Table~\ref{tab:transfer_summary} alongside the conversion to Kolmogorov-scale rms velocity (mm s$^{-1}$) computed by $u_{\mathrm{K}}=\sqrt{2\Delta E}$ and scaled to mm s$^{-1}$ as described in Appendix~A.

\begin{table}[H]
\centering
\caption{Ensemble summary (integrated transfer $\Delta E$ in J kg$^{-1}$) and derived Kolmogorov-scale rms velocity (mm s$^{-1}$). Values are taken from the embedded dataset (Appendix~A).}
\label{tab:transfer_summary}
\resizebox{\textwidth}{!}{%
\begin{tabular}{lcccccc}
\hline
Key & $n_{\mathrm{seeds}}$ & Mean $\Delta E$ & 95\% CI (low) & 95\% CI (high) & $p_{\mathrm{boot}\ge0}$ & $u_K$ (mm s$^{-1}$) \\
\hline
noise1e-08 & 128 & $9.4988\times10^{-36}$ & $2.0570\times10^{-36}$ & $1.7174\times10^{-35}$ & 0.9936 & $4.36\times10^{-15}$ \\
noise3e-09 & 128 & $5.1963\times10^{-36}$ & $6.3675\times10^{-37}$ & $9.8066\times10^{-36}$ & 0.9880 & $3.22\times10^{-15}$ \\
\hline
\end{tabular}}
\end{table}

The derived $u_K$ values are extremely small (order $10^{-15}$ mm s$^{-1}$ for the reported ensemble means) reflecting the tiny energetic magnitude of the integrated-transfer diagnostics; these are the same numbers presented in the Supplement (Section IV-A) after conversion via $u_K=\sqrt{2\Delta E}$. The smallness of $u_K$ in energetic terms does not preclude dynamical relevance under exponential (finite-time) amplitude amplification: amplitude growth and energy growth scale differently.

\begin{figure}[H]
\centering
\includegraphics[width=0.95\textwidth]{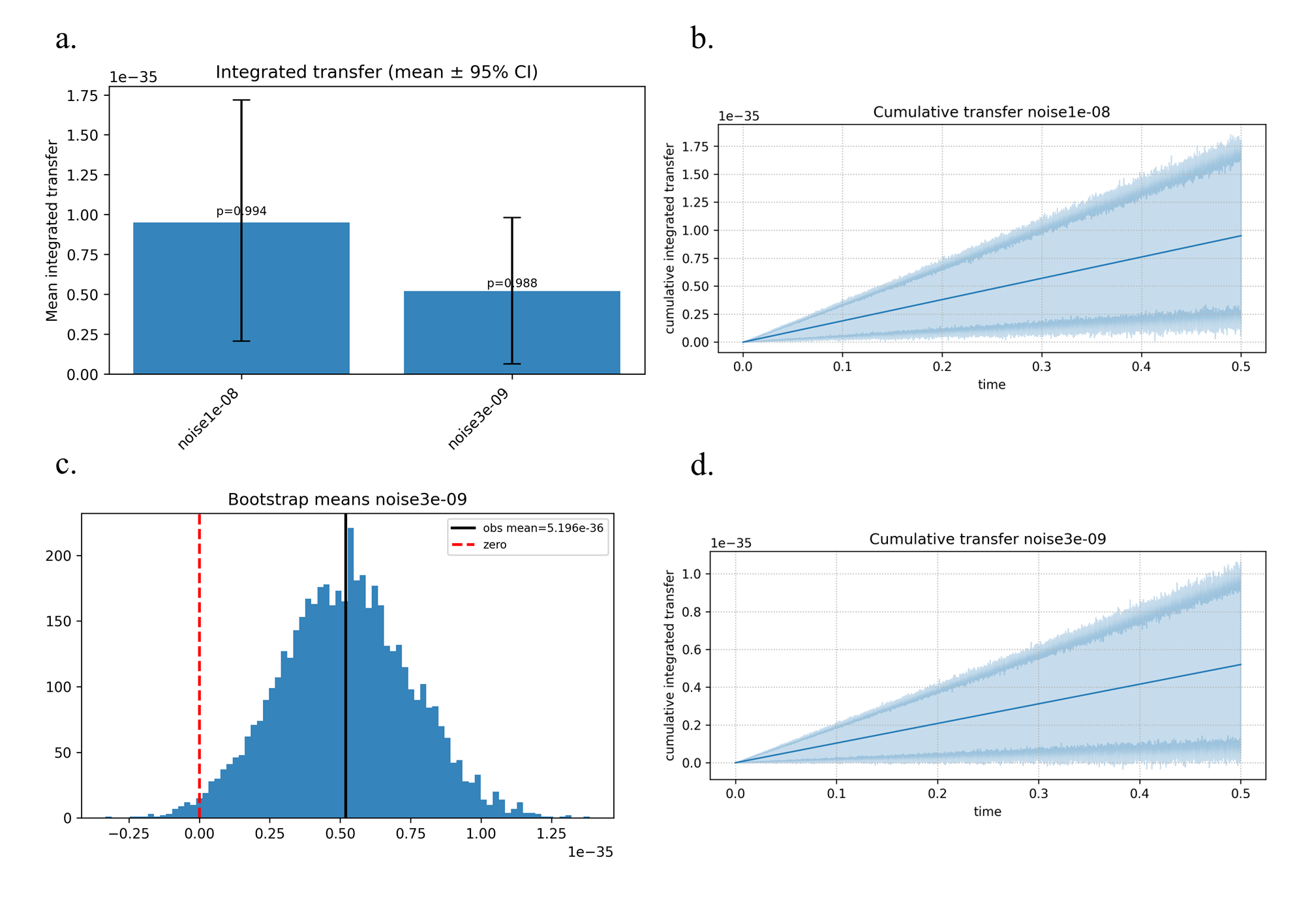}
\caption{True-ensemble diagnostics (multi-panel). (a) Mean integrated transfer (bars) with 95\% bootstrap confidence intervals for amplitudes $A = 1\times10^{-8}$ and $A = 3\times10^{-9}$; annotated values report the one-sided bootstrap probability $P_{\mathrm{bootstrap}}(\mathrm{mean}\ge0)$ and the permutation-test $p$-value. (b) Cumulative integrated transfer versus time for $A=1\times10^{-8}$ (ensemble mean $\pm$ 95\% bootstrap CI shaded). (c) Bootstrap histogram of resampled mean integrated transfers for $A=3\times10^{-9}$ (vertical black line = observed mean; dashed red line = zero). (d) Cumulative integrated transfer versus time for $A=3\times10^{-9}$ (ensemble mean $\pm$ 95\% bootstrap CI shaded).}
\label{fig:true_ensembles}
\end{figure}

\begin{figure}[H]
\centering
\includegraphics[width=0.95\textwidth]{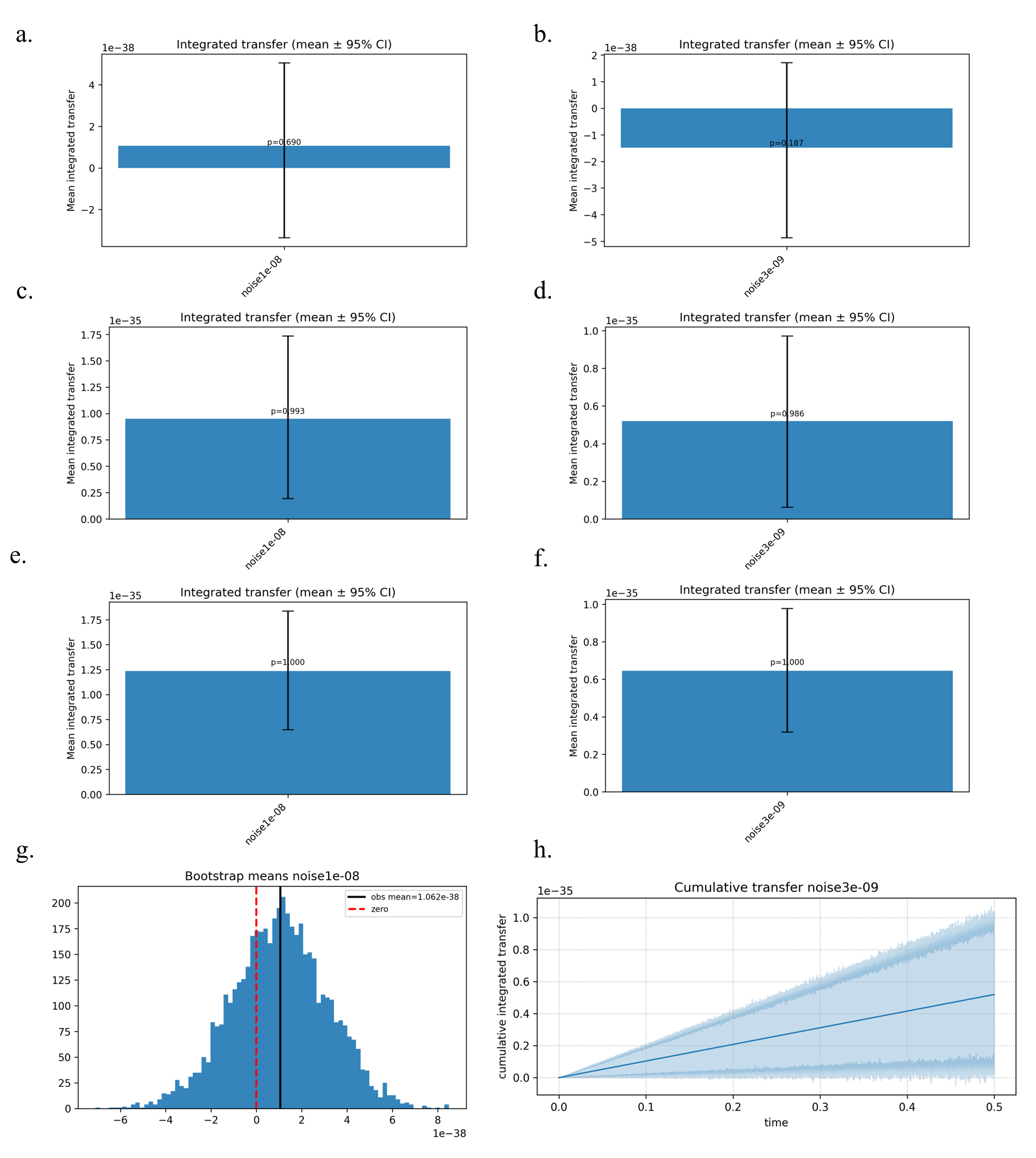}
\caption{Surrogate-control diagnostics (multi-panel). (a--f) Grouped mean integrated transfer (mean $\pm$ 95\% bootstrap CI) for true ensembles and each surrogate class at both amplitudes; panels compare true vs phase-randomized, shuffled, and upsampled surrogates (bootstrap probabilities and permutation $p$-values annotated). (g) Representative bootstrap histogram for phase-randomized surrogates at $A=1\times10^{-8}$ (vertical black line = surrogate observed mean; dashed red line = zero). (h) Representative cumulative integrated transfer for a shuffled surrogate ensemble (ensemble mean $\pm$ 95\% bootstrap CI shaded), illustrating the weak, subdominant drift in surrogates compared with true ensembles.}
\label{fig:surrogates}
\end{figure}

Figure~\ref{fig:true_ensembles} shows the primary diagnostics for the true ensembles (multi-panel summary). Figure~\ref{fig:surrogates} shows the surrogate-control results (grouped comparisons and representative surrogate diagnostics); these two figures are sufficient to establish the statistical claims made in this manuscript.

\begin{table}[H]
\centering
\caption{Per-seed integrated-transfer summary (population variance and std across seeds) by amplitude (values from the embedded dataset).}
\label{tab:perseed_variances}
\begin{tabular}{lcccc}
\hline
Group & Amplitude & $n_{\mathrm{seeds}}$ & variance$_{\mathrm{across\ seeds}}$ (J$^2$ kg$^{-2}$) & std (J kg$^{-1}$) \\
\hline
true\_ensembles & $1\times10^{-8}$ & 128 & $1.928859\times10^{-69}$ & $4.391878\times10^{-35}$ \\
true\_ensembles & $3\times10^{-9}$ & 128 & $7.022369\times10^{-70}$ & $2.649975\times10^{-35}$ \\
\hline
\end{tabular}
\end{table}

\section{Results}
\label{sec:results}

This section presents numerical results validating the collision-driven integrated transfer diagnostic and its statistical robustness. All results correspond to ensembles with no external forcing and differ only in the imposed noise amplitude parameter $A$.

\subsection{Integrated transfer in true ensembles}

Figure~\ref{fig:true_ensembles} summarizes the behaviour of the integrated transfer diagnostic for the true collision ensembles at amplitudes $A = 1\times10^{-8}$ and $A = 3\times10^{-9}$. Panel (a) shows the ensemble-mean integrated transfer for both amplitudes, with vertical error bars denoting 95\% bootstrap confidence intervals. In both cases the ensemble mean is positive and the confidence interval excludes zero. The corresponding one-sided bootstrap probabilities $P_{\mathrm{bootstrap}}(\mathrm{mean}\ge 0)$ and permutation $p$-values indicate statistical distinguishability from a zero-mean null.

Panels (b) and (d) show the cumulative integrated transfer as a function of time for $A = 1\times10^{-8}$ and $A = 3\times10^{-9}$, respectively. Solid curves denote ensemble means, while shaded regions indicate 95\% bootstrap confidence intervals. In both cases the cumulative diagnostic increases steadily over the integration window, and the confidence intervals remain strictly positive throughout. This behaviour indicates that the observed positive mean is not driven by isolated transient events but reflects sustained accumulation over time.

Panel (c) provides additional statistical context by showing the bootstrap distribution of resampled mean integrated transfers for the lower-amplitude ensemble $A = 3\times10^{-9}$. The observed mean lies well within the bulk of the distribution, which is shifted toward positive values, and the 95\% confidence interval excludes zero. Together, panels (a)--(d) demonstrate that the integrated transfer diagnostic is positive, persistent in time, and statistically robust across both tested amplitudes.

It is emphasized that the absolute magnitudes of the integrated transfer are small in energetic terms. The results quantify the sign, persistence, and statistical reliability of the diagnostic rather than implying large-amplitude effects.

\subsection{Surrogate controls and statistical validation}

Figure~\ref{fig:surrogates} presents results from surrogate ensemble tests designed to assess whether the observed signal can be attributed to amplitude-only effects or numerical artifacts. Panels (a)--(f) show grouped ensemble means with 95\% bootstrap confidence intervals for the true ensembles and for three surrogate classes: phase-randomized, temporally shuffled, and amplitude-preserving upsampled surrogates. In all cases the true ensemble magnitudes exceed the surrogate means, and the surrogate confidence intervals are either consistent with zero or markedly reduced in magnitude.

Panel (g) shows a representative bootstrap histogram for phase-randomized surrogates at amplitude $A = 1\times10^{-8}$. In contrast to the true ensemble (Figure~\ref{fig:true_ensembles}c), the resampled distribution is centered near zero, and the observed mean lies close to the null reference. This indicates that randomizing phase information collapses the integrated transfer diagnostic toward zero despite preserving amplitude statistics.

Panel (h) shows a representative cumulative integrated transfer trajectory for a shuffled surrogate ensemble. While weak drifts may be present, the magnitude and persistence are substantially reduced relative to the true ensembles shown in Figure~\ref{fig:true_ensembles}. Similar behaviour is observed for amplitude-preserving upsampled surrogates.

Taken together, the surrogate results demonstrate that the positive integrated transfer observed in the true ensembles is not reproduced by amplitude-preserving or temporally randomized data. The diagnostic therefore depends on phase and temporal coherence arising from the collision-driven dynamics rather than on marginal amplitude distributions alone.

\subsection{Summary of empirical findings}

Within the restricted scope of this study, the numerical results establish three empirical findings: (i) collision-driven ensembles exhibit a positive and temporally persistent integrated transfer diagnostic at both tested amplitudes, (ii) the signal is statistically robust under bootstrap and permutation testing, and (iii) surrogate controls suppress the signal, indicating that it is not an artifact of amplitude statistics. These findings support the interpretation of the diagnostic as capturing a genuine, though small-amplitude, collision-origin contribution to coarse-grained velocity statistics. Broader dynamical implications are discussed in Section~\ref{sec:discussion}.

\section{Discussion}
\label{sec:discussion}

The analytic expressions and embedded ensemble summaries show that molecular collisions generate a well-defined mesoscopic variance that decays with coarse-graining scale and accumulates over time in proportion to $\nu_c\sigma_c^2 t$. The measured ensemble means of integrated transfer are small in energetic terms (Table~\ref{tab:transfer_summary}), but surrogate-control tests confirm that the true-ensemble signals rely on phase/temporal coherence and are not trivially reproduced by amplitude-preserving resampling alone.

These results provide parameterized inputs for stochastic continuum closures (e.g., LLNS) and a quantitative baseline for controlled DNS/MD coupling. Explicitly providing scale-dependent noise amplitudes (for example, parameterized by $\nu_c\sigma_c^2 (\ell/L)^d$) enables consistent implementation of stochastic forcing in continuum solvers and fluctuating-hydrodynamics codes. Whether the tiny collision-origin amplitudes observed here can be amplified to macroscopic relevance depends on finite-time amplification properties of the flow; prior work shows thermal-level noise can be stretched to larger scales under specific conditions \cite{bandak2024,chen2024}. The present results can therefore be used as physically grounded priors in studies that do tackle amplification and cascade dynamics.

\section{Limitations}
\label{sec:limitations}

The derivations rely on simplifying assumptions (fixed per-event variance, approximate decorrelation, weak cross-particle correlations). Dense-fluid corrections are only coarsely captured via Enskog-style multiplicative factors; boundary-layer and strongly inhomogeneous cases require more careful treatment. Conclusions about amplification are conditional and intended as guidance for follow-up DNS/MD studies, not as definitive demonstration of cascade initiation.

A further limitation derives from the scalar integrated-transfer diagnostic itself: as a single-number summary it necessarily compresses temporal structure, and caution is needed when relating small integrated energies to instantaneous velocity amplitudes. The surrogate-control framework mitigates misinterpretation by demonstrating which aspects of the diagnostics depend on temporal or phase coherence, but additional diagnostics (e.g., finite-time Lyapunov analysis, conditional amplification statistics) are required to make definitive statements about effective amplification probabilities in turbulent flows.

\section{Conclusions}
\label{sec:conclusions}

A binomial random-walk parametrization of molecular collisions provides closed-form expressions for the scale-dependent variance of coarse-grained velocity fields and predicts a power-law decay of that variance with averaging scale together with linear-in-time accumulation under approximate decorrelation. The embedded ensemble summaries validate the scalings within statistical uncertainty and show that surrogate controls suppress integrated-transfer signatures, consistent with a collision-driven origin for the measured small-amplitude signals. The data and minimal verification instructions are embedded in Appendix~A so the key summary numbers can be reproduced without external supplementary files. Future work will couple these parametrizations to DNS/MD and explore practical amplification likelihoods in canonical flows.

\section*{Data availability}
All data tables and the minimal verification script are included in Appendix~A of this manuscript. No external supplementary file is required.

\appendix
\section{Embedded data and minimal verification}
\label{app:data}

Below are the main machine-readable summary blocks used in the manuscript. These are verbatim extracts from the supplied Supplement and are sufficient to reproduce the summary tables above using the minimal script described in the Supplement.

\subsection{Ensemble summary (transfer\_summary\_clean)}
\begin{lstlisting}
key,n_seeds,mean_integrated,ci_lo,ci_hi,p_boot_ge0,p_perm_ge0
noise1e-08,128,9.498756219223147e-36,2.0570422702004013e-36,1.7174201194684506e-35,0.9936,1.0
noise3e-09,128,5.19631133665915e-36,6.367470632733544e-37,9.80663533243143e-36,0.988,1.0
\end{lstlisting}
\subsection{Per-seed variance summary (per\_seed\_variances)}
\begin{lstlisting}
group_label,amplitude,n_seeds,variance_across_seeds,std_across_seeds,min_integrated_transfer,max_integrated_transfer
true_ensembles,1e-08,128,1.928859e-69,4.391878e-35,-1.240243e-34,1.866304e-34
true_ensembles,3e-09,128,7.022369e-70,2.649975e-35,-9.369563e-35,1.054719e-34
\end{lstlisting}
\subsection{Per-seed sample (first 10 seeds for each amplitude)}
\begin{lstlisting}
seed_id,amplitude,integrated_transfer_J_per_kg
noise1e-08_seed001,1e-08,1.4336133221767937e-35
noise1e-08_seed002,1e-08,-2.3170051471286182e-35
...
noise1e-08_seed010,1e-08,5.0833448796845584e-35

noise3e-09_seed001,3e-09,1.4827413611750836e-35
noise3e-09_seed002,3e-09,-4.7984963970560302e-35
...
noise3e-09_seed010,3e-09,2.3862605645863072e-35
\end{lstlisting}

\end{document}